\documentclass[aps,pre,twocolumn,showpacs,floatfix]{revtex4}
\usepackage{color} % for use with pdflatex
\usepackage{latexsym,hyperref,amssymb,graphicx,dcolumn,bm}
\newcommand{\thalf}{{\textstyle{\frac{1}{2}}}}

\newcommand{\beq}{\begin{equation}}
\newcommand{\eeq}{\end{equation}}
\newcommand{\bea}{\begin{eqnarray}}
\newcommand{\eea}{\end{eqnarray}}
\newcommand{\bean}{\begin{eqnarray*}}
\newcommand{\eean}{\end{eqnarray*}}
\newcommand{\bei}{\begin{itemize}}
\newcommand{\eei}{\end{itemize}}
\newcommand{\ben}{\begin{enumeration}}
\newcommand{\een}{\end{enumeration}}

\definecolor{navyblue}{rgb}{.05,0,.55}
\newcommand{\tcr}[1]{\textcolor{red}{#1}}

\newcommand{\tcnb}[1]{\textcolor{navyblue}{#1}}

\newcommand{\tcm}[1]{\textcolor{magenta}{#1}}

\begin{document}

\title{Parametrization of the Hybrid Potential for Pairs of Neutral Atoms}

\author{Kevin Cahill}
\email{cahill@unm.edu}
\affiliation{Department of Physics and Astronomy, 
University of New Mexico, Albuquerque, NM 87131}
\date{\today}
\begin{abstract}
The hybrid form
is a combination
of the Rydberg potential 
and the London inverse-sixth-power energy.
It is accurate at all relevant distance scales
and simple enough for use in all-atom simulations 
of biomolecules.
One may compute the parameters
of the hybrid potential for the ground state of a pair of
neutral atoms from their internuclear separation,
the depth and curvature 
of their potential at its minimum,
and from their 
van der Waals coefficient of dispersion \(C_6\)\@.
\end{abstract}


\maketitle
\section{The Hybrid Form\label{bad forms}}
Commercial molecular-modeling codes use
pair potentials chosen more for speed
than for accuracy.
They use the harmonic potential
\beq
V_h(r) = -E_0 + \frac{k}{2} \, (r - r_0)^2
\label {harmonic}
\eeq
for covalently bonded pairs
of neutral atoms and the
Lennard-Jones potential~\cite{Lennard-Jones1931}
\beq
V_{LJ}(r) = E_0 \left[ \left( \frac{r_0}{r} \right)^{12}
- 2 \, \left( \frac{r_0}{r} \right)^6 \right]
\label {LJ}
\eeq
for unbonded pairs.
The recently introduced
hybrid potential~\cite{CahillPar2004}
\beq
V(r) = a e^{-b \, r}\,( 1 - c \, r )
- \frac{C_6}{r^6 + d \, r^{-6}}
\label{hybrid}
\eeq
is nearly as fast and much more accurate.
It is fast enough for use in all-atom
simulations of biomolecules and accurate
at all biologically relevant distance scales
when its parameters are suitably chosen.
How to choose them is the focus of this paper.
Formulas are derived for
\(a\), \(b\), \(c\), \& \(d\) in terms of
the internuclear separation \(r_0\),
the depth \(E_0\) and curvature \(k\) 
of the potential at its minimum,
and the van der Waals coefficient \(C_6\) 
of the pair.
The hybrid potential therefore is applicable
to pairs of neutral atoms for which
no empirical potential is available.
Its eventual incorporation into
\textsc{tinker}~\cite{Ponder2003},
Amber~\cite{Amber71995}, and
other such codes
is a goal of this work.
\par
The hybrid potential is a combination of the Rydberg formula
used in spectroscopy
and the London formula for pairs of atoms.
The terms involving  
\(a\), \(b\), and \(c\)
were proposed by Rydberg
to incorporate spectroscopic data,
but were largely ignored until recently.
The constant \(C_6\)
is the coefficient of the London tail.
The new term \( d \, r^{-6} \) cures
the London singularity.
As \(r \to 0\),
\( V(r) \to a\), finite;
as \(r \to \infty\), \(V(r)\)
approaches the London term,
\(V(r) \to \mbox{} - C_6/r^6\)\@.
In a perturbative analysis~\cite{CahillPar2003},
the \(a, b, c\) terms
arise in first order, and the \(C_6\) term
in second order.
\par
When fitted to spectroscopically
determined potentials for 
the ground states of
H\(_2\), N\(_2\), O\(_2\),
Ar--Ar, and Kr--Kr,
the hybrid form is four orders
of magnitude more accurate
than the harmonic 
and Lennard-Jones potentials
and five times more accurate than 
the Morse~\cite{Morse1929}, Varnshi~\cite{Varnshi1957},
and Hulburt-Hirschfelder~\cite{Hulbert1961} 
potentials~\cite{CahillPar2004}.
It also yields accurate second virial coefficients
and heats of vaporization~\cite{CahillPar2004}\@.
Its simplicity recommends it as a teaching tool
and as a practical form for computation.
\par
How does one find the parameters
\(a\), \(b\), \(c\), \& \(d\)
when an empirical potential is not
available?
For many pairs of neutral atoms,
spectroscopists have measured  
the internuclear separation \(r_0\),
the well depth \(E_0 = |V(r_0)|\) and 
curvature \(k = V^{''}(r_0)\)
at the minimum of the potential, and the
London coefficient \(C_6\)\@.
These input parameters are discussed
in Sec.~\ref{inputPars} with an emphasis on
the curvature \(k\) and its relation
to the vibrational
frequency of the ground state and
to the energy \(D_0\) needed to
dissociate the ground state.
Section~\ref{rule of thumb} develops
a rule of thumb for the parameter \(d\)
to which the hybrid form is 
relatively insensitive.
Section~\ref{abc} derives formulas
for the hybrid parameters \(a\), \(b\), \& \(c\)
in terms of \(r_0\),
\(E_0\), \(k\), \(C_6\), \& \(d\)
and plots the resulting hybrid potentials
for the 11 pairs
H\(_2\), N\(_2\), O\(_2\), NO, OH, I\(_2\), 
Li\(_2\), Na\(_2\), K\(_2\),  Ar--Ar, \& Kr--Kr\@.
\section{Input Parameters\label{inputPars}}
In addition to the well depth
\(E_0 = |V(r_0)|\), the internuclear separation
\(r_0\) at the minimum,
and the London or van der Waals coefficient \(C_6\),
the curvature \(k = V^{''}(r_0)\) of the potential
at \(r_0\) often is available either directly
or in the guise of the energy 
of dissociation \(D_0\)\@.  
\par
Near \(r_0\),
the motion of the internuclear separation \(r\)
is described by the lagrangian
\beq
L = \thalf \, \mu \, {\dot r}^2 - \thalf \, k \, (r - r_0)^2
\label {Lr0}
\eeq
in which \(\mu = m_1 m_2/(m_1 + m_2 ) \)
is the reduced mass of the two atoms
of mass \(m_1\)
and \(m_2\)\@.
To lowest order in \(r - r_0\),
the ground state then has energy
\beq
E_g = \thalf \hbar \omega - E_0
\label {E_g}
\eeq
in which  \(\omega\) is the (angular) 
vibrational frequency of the ground state.
The curvature \(k\) is related to \(\omega\) by
\beq
k = \mu \omega^2.
\label {k}
\eeq 
\par
In the chemical-physics literature,
the depth \(E_0\) and 
separation \(r_0\)
are labeled \(D_e\) and \(r_e\), and 
the angular frequency \(\omega = 2 \pi\nu\)
is expressed as a frequency \(\omega_e\)
in inverse centimeters:
\(\omega_e = \omega/(2 \pi c)\)
where \(c\) is the speed of light.
In these terms and in cgs units, 
\(k = 4 \pi^2 c^2 \mu \, \omega_e^2\)\@.
The energy of dissociation
is \( D_0 = -E_g = E_0 - \hbar \omega/2 \),
and so the curvature \(k = V^{''}(r_0)\) is
related to the difference between it and the well depth
\(E_0 = D_e\) by
\beq
k = 4 \mu (E_0 - D_0)^2/\hbar^2 = 4 \mu (D_e - D_0)^2/\hbar^2 .
\label {k from E_0 - D_0}
\eeq
\par
The values of the input parameters
are listed in Table~\ref{tab:table1}\@.
The internuclear separation \(r_0\) at the minimum 
of the potential well \(V(r)\) and 
its depth \(E_0 = V(r_0)\)
are from~\cite{Vanderslice1962} 
for H\(_2\), NO, OH, \& I\(_2\),
and from~\cite{Krupenie1972} for O\(_2\)\@.
The curvature \(k = V^{''}(r_0)\) of the potential
at its minimum is from~\cite{Vanderslice1962}
for H\(_2\), O\(_2\), NO, OH, \& I\(_2\)\@.
The van der Waals coefficient
\(C_6\) is from~\cite{Bromley2005} for H\(_2\);
from~\cite{Zeiss&Meath1977}
for NO; from~\cite{Varandas1995} for OH; and 
from~\cite{Dalgarno2004}
for O\(_2\), I\(_2\),
Ar--Ar, \& Kr--Kr\@.
The values of \(r_0\), \(E_0\),
\& \(k\) 
for Ar--Ar and Kr--Kr respectively are
from~\cite{Aziz1993} and~\cite{Dham1989}\@. 
The values of \(r_0\), \(E_0\), \(k\), \& \(C_6\)
for N\(_2\), Li\(_2\), Na\(_2\), \& K\(_2\) 
respectively are from~\cite{Jary2006};
\cite{Melville2006};
\cite{Dalgarno2004}, \cite{Tiemann2000}, \&
\cite{Johnson1999}; and
\cite{Dalgarno2004},
\cite{Johnson1999}, 
\cite{Amiot1995},
\& \cite{Amiot1991}\@.
\par
\section{A Rule of Thumb\label{rule of thumb}}
The \(d\)-term of
the hybrid form is a trick to avoid
the \(1/r^6\) singularity of the London term.
It has no other justification. 
Luckily, the quality
of the fit is not sensitive to the precise value
of the parameter \(d\),
and so we need only a good rule of thumb
for that parameter.
To develop an approximate formula that 
estimates this parameter, we 
first find values of \(a\), \(b\), \(c\), \& \(d\)
that provide good fits of the hybrid form \(V(r)\) 
to empirical potentials 
for the 11 pairs
H\(_2\), N\(_2\), O\(_2\), NO, OH, I\(_2\), 
Li\(_2\), Na\(_2\), K\(_2\),
Ar--Ar, \& Kr--Kr
of neutral atoms.

\begin{table}
\caption{\label{tab:table1}The depth \(E_0 = V(r_0)\) 
of the potential well \(V(r)\), 
the internuclear separation \(r_0\)
and the curvature \(k = V^{\prime\prime}(r_0)\)
at the minimum \(r = r_0\), and the van der Waals coefficient
\(C_6\) for 11 pairs of neutral atoms.}
\begin{ruledtabular}
\begin{tabular}{|c|c|c|c|c|} 
\ & \(E_0\) (eV) & \(r_0\) (\AA) 
& \(k\) (eV\,\AA\(^{-2}\)) & \(C_6\) (eV\,\AA\(^6\)) \\ 
\hline 
H\(_2\) & 4.7467 & 0.7417 & 35.8861& 3.88338  \\ \hline
\(^{14}\)N\(_2\) & 9.8995 & 1.09768\footnotemark[1] & 143.2245 & 14.382  \\ \hline
O\(_2\) & 5.2136 & 1.2075 &  73.4726 & 9.3215
\\ \hline
NO & 6.609 & 1.1590 &  99.5676 & 11.245 \\ \hline
OH & 4.624 & 0.9707 &  48.6196 &  6.854 \\ \hline
I\(_2\) & 1.5571 & 2.668 &  10.7378 & 230.05 \\ \hline
\(^7\)Li\(_2\) & 1.0559 & 2.6730 & 1.5752 & 829.33 \\ \hline
Na\(_2\) & 0.74664 &  3.0786 & 1.0706 & 929.76 \\ \hline
K\(_2\) & 0.55183 & 3.9243 & 0.61375 & 2328.6 \\ \hline 
Ar\(_2\) & 0.01234 & 3.757   &  0.0691 & 38.4213  \\ \hline
Kr\(_2\) & 0.01735 & 4.017   &  0.0896 & 77.6791  \\ \hline
\end{tabular}
\end{ruledtabular}
\footnotetext[1]{Extra digits are included to avoid round-off errors.}
\end{table}

\par
Empirical potentials obtained from spectroscopic 
data~\cite{Weissman1963,Krupenie1977,Krupenie1972}
by the RKR 
(Rydberg~\cite{Rydberg1931}, Klein~\cite{Klein1932},
Rees~\cite{Rees1947}) method are available
from~\cite{Vanderslice1962} 
for H\(_2\), NO, \& I\(_2\); 
from~\cite{Groenenboom2007} for OH;
from~\cite{Jary2006} for N\(_2\);
from~\cite{Krupenie1972} for O\(_2\);
from~\cite{Melville2006} for Li\(_2\);
from~\cite{Tiemann2000} for Na\(_2\);
from~\cite{Stwalley1998} for K\(_2\);
from~\cite{Aziz1993} for Ar--Ar; and
from~\cite{Dham1989} for Kr--Kr\@.
\par
\begin{figure}
\centering
\input {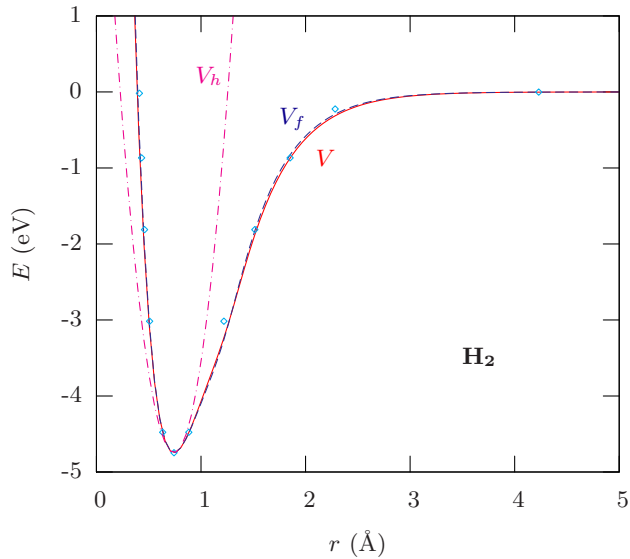}
\caption{The hybrid form \(\tcr{V}\) 
with the calculated coefficients 
(\(a\), \(b\), \(c\), \& \(d\), Table~\ref{tab:table3}) 
(solid, red) fits
the RKR spectral points for 
the ground state of
molecular hydrogen (diamonds, cyan)
and gives the correct London tail for \(r > 3\) \AA\
nearly as well as does the hybrid form \(\tcnb{V_f}\) 
with the fitted coefficients 
(\(a\), \(b\), \(c\), \& \(d\), Table~\ref{tab:table2})
(dashes, dark blue)\@.
In all the figures,
the values of \(C_6\) used 
in \(\tcr{V}\) and \(\tcnb{V_f}\)
are from Table~\ref{tab:table1}\@. 
The harmonic form
\(\tcm{V_h}\) (dot-dash, magenta) fits
only near the minimum.}
\label{rkrH2}
\end{figure}
\begin{figure}
\centering
\input {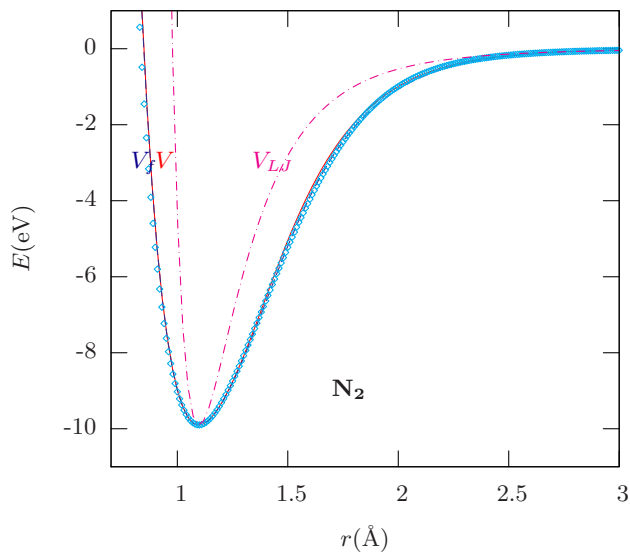}
\caption{The hybrid form \(\tcr{V}\) 
with the calculated coefficients 
(\(a\), \(b\), \(c\), \& \(d\), Table~\ref{tab:table3})
(solid, red) fits
the RKR spectral points for 
the ground state of
molecular nitrogen (diamonds, cyan)
nearly as well as does the hybrid form \(\tcnb{V_f}\) 
with the fitted coefficients 
(\(a\), \(b\), \(c\), \& \(d\), Table~\ref{tab:table2})
(dashes, dark blue)\@.
The Lennard-Jones form \(\tcm{V_{LJ}}\) (dot-dash, magenta)
fits only near the minimum.}
\label{rkrN2}
\end{figure}
\begin{figure}
\centering
\input {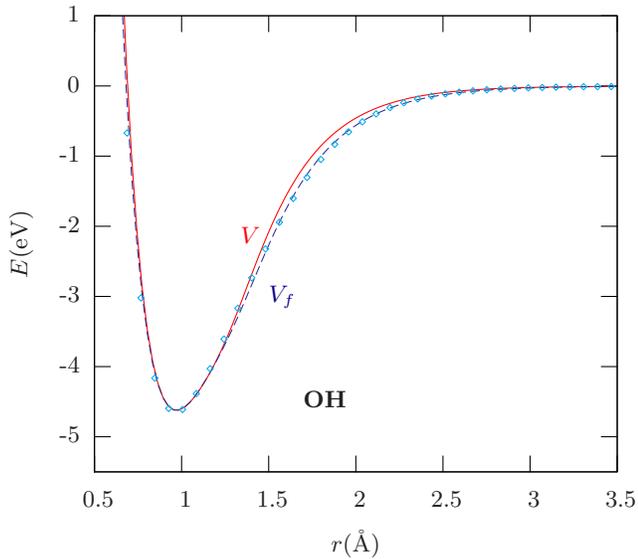}
\caption{The hybrid form \(\tcr{V}\) 
with the calculated coefficients 
(\(a\), \(b\), \(c\), \& \(d\), Table~\ref{tab:table3})
(solid, red) fits
the RKR spectral points for 
the ground state of the
hydroxyl radical (diamonds, cyan)
slightly less well than the hybrid form \(\tcnb{V_f}\) 
with the fitted coefficients 
(\(a\), \(b\), \(c\), \& \(d\), Table~\ref{tab:table2})
(dashes, dark blue)\@.}
\label{rkrOH}
\end{figure}
\begin{figure}
\centering
\input {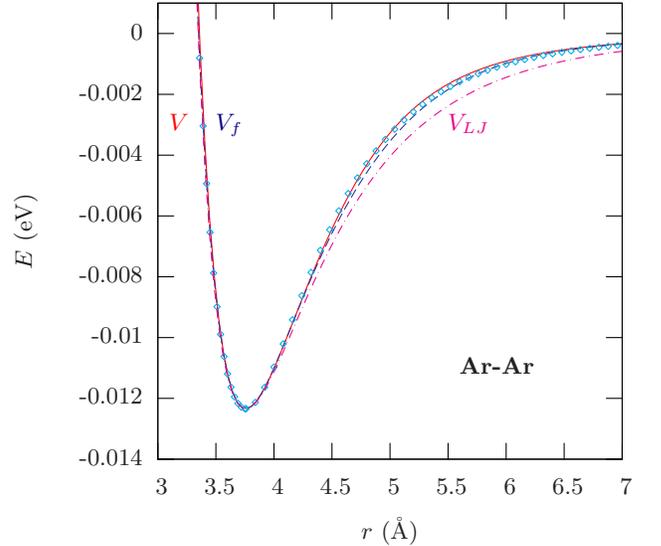}
\caption{The hybrid form \(\tcr{V}\) 
with the calculated coefficients 
(\(a\), \(b\), \(c\), \& \(d\), Table~\ref{tab:table3})
(solid, red) fits
the RKR spectral points for 
the ground state of 
the argon dimer (diamonds, cyan)
nearly as well as does the hybrid form \(\tcnb{V_f}\) 
with the fitted coefficients 
(\(a\), \(b\), \(c\), \& \(d\), Table~\ref{tab:table2})
(dashes, dark blue)\@.
The Lennard-Jones form \(\tcm{V_{LJ}}\) 
fitted to the minimum (dot-dash, magenta)
is too low for \(r > 4\) \AA\@.}
\label{rkrAr2}
\end{figure}

Section~\ref{abc} contains formulas
(\ref{ais}, \ref{bis}, \ref{cis}, \& \ref{minus sign})
for \(a,\, b,\, \&\, c\) in terms of \(E_0,\, r_0,\, k,\, C_6,\, \& \, d\)
that ensure that the hybrid form \(V(r)\)
goes through the minimum \((r_0, -E_0)\)
with curvature \(k\)\@.
For each pair of atoms,
I found the value
of \(d\) that best fits the hybrid form \(V(r)\)
to the empirical potential of the pair
for \(r \ge f \, r_0\)
in which \(0 \le f \le 1\).
The resulting values of \(a\), \(b\), \(c\), \& \(d\)
are listed in Table~\ref{tab:table2}
along with the fractions \(f\) and the root-mean-square
(rms) errors \(\Delta V(r)\) for \(r \ge f \, r_0\)\@.
Figures~\ref{rkrH2}--\ref{rkrKr2} show 
that the fitted hybrid forms \(\tcnb{V_f}\) 
(dashes, dark blue)
nicely follow the points 
(diamonds, cyan) of the empirical potentials
for the 11 pairs of neutral atoms
H\(_2\), N\(_2\), O\(_2\), NO, OH, I\(_2\), 
Li\(_2\), Na\(_2\), K\(_2\),
Ar--Ar, \& Kr--Kr\@.  
The fitted \(\tcnb{V_f(r)}\)'s go
through the empirical minima with the
right curvatures and closely trace the
empirical potentials, at least for \(r \ge f r_0\)\@.
Figure~\ref{rkrH2}
for molecular hydrogen
adds
the harmonic potential \(\tcm{V_h}\)
(dot-dash, magenta),
which fits only near
the minimum at \(r = r_0\)\@.
Figure~\ref{rkrN2}
for molecular nitrogen and 
Fig.~\ref{rkrAr2} for a pair of argon atoms
include 
the Lennard-Jones potential 
\(\tcm{V_{LJ}}\) (dot-dash, magenta),
which is accurate only near \(r_0\)\@.
\begin{table}
\caption{\label{tab:table2} The values of the coefficients
\(a\), \(b\), \(c\), \& \(d\) that fit 
the hybrid potential \(V(r)\) (Eq.~(\ref{hybrid})) 
to RKR data for H\(_2\), N\(_2\), O\(_2\), NO, OH, I\(_2\),
Li\(_2\), Na\(_2\), \& K\(_2\)
and to empirical potentials for Ar--Ar \& Kr--Kr
while respecting Eqs.~(\ref{ais}, \ref{bis}, \ref{cis}, \& \ref{minus sign})\@.
The rms errors \(\Delta V(r)\) for \(r \ge f \, r_0\) are also listed.}
\begin{ruledtabular}
\begin{tabular}{|c|c|c|c|c|c|}  
\  & \(a\) (eV) & \(b\) (\AA\(^{-1}\)) & \(c\) (\AA\(^{-1}\)) 
& \(d\) (\AA\(^{12}\)) & f,\,\, \(\Delta V\!\) (eV)\\
\hline
H\(_2\) & 47.796 & 2.9632 & 2.5406 & 12.2  & 0.68, 0.086 \\ \hline 
N\(_2\) & 3752.644 & 4.3533 & 1.1777 & 34.8 & 0.81, 0.088 \\ \hline
O\(_2\) & 2901.580 & 4.2173 & 1.0510 & 59.8 & 0.84, 0.039\\ \hline
NO      & 3809.497 & 4.4196 & 1.0943 &  47.0 & 0.81, 0.077\\ \hline
OH      &  377.804 & 3.6909 & 1.4668 & 32.5 & 0.80, 0.023 \\ \hline
I\(_2\) & 14361.15 & 2.8013 & 0.4351 &  2.08e5\footnotemark[2] & 0.88, 0.039 \\ \hline
Li\(_2\)& 199.481 & 1.6200 & 0.5101 & 2.85e6 & 0.86, 0.016 \\ \hline
Na\(_2\)& 231.900  & 1.5311 & 0.4292 &  9.40e6  & 0.81, 0.009 \\ \hline
K\(_2\) & 325.051 & 1.3409 & 0.3269 &  9.94e7  & 0.85, 0.008 \\ \hline
Ar\(_2\)& 1987.943 & 2.6517 & 0.2978 & 8.10e7  & 0.70, 0.00009 \\ \hline
Kr\(_2\)& 1875.462 & 2.3661 & 0.2789 & 5.72e8  & 0.90, 0.00018\\ \hline
\end{tabular}
\end{ruledtabular}
\footnotetext[2]{2.083e5 = 2.083\(\times 10^5\)\@.}
\end{table}
\begin{table}
\caption{\label{tab:table3}Values of the coefficients
\(a\), \(b\), \(c\), and \(d\) obtained by
Eqs.~(\ref{d emp}, \ref{ais}, \ref{bis}, \ref{cis}, \& \ref{minus sign})
from the values of \(E_0\), \(r_0\), \(k\), 
and \(C_6\) of Table~\ref{tab:table1}\@.}
\begin{ruledtabular}
\begin{tabular}{|c|c|c|c|c|c|}  
\  &\(a\) (eV) & \(b\) (\AA\(^{-1}\)) 
& \(c\) (\AA\(^{-1}\)) & \(d\) (\AA\(^{12}\)) & f, \(\Delta V\) \\ \hline 
H\(_2\)  &     45.01 &  2.907 &  2.5663 &  16.7 & 0.68, 0.087\\ \hline
N\(_2\)  &   4059.02 &  4.435 &  1.1762 &  27.7 & 0.81, 0.111\\ \hline
O\(_2\)  &   2868.48 &  4.246 &  1.0539 &  40.7 & 0.84, 0.068\\ \hline
NO       &   4040.42 &  4.496 &  1.0946 &  34.0 & 0.81, 0.156\\ \hline
OH       &    491.96 &  3.942 &  1.4478 &  19.8 & 0.81, 0.053\\ \hline
I\(_2\)  &  16125.67 &  2.832 &  0.4350 &  2.79e5 & 0.88, 0.041\\ \hline
Li\(_2\) &    148.93 &  1.516 &  0.5161 &  4.01e6 & 0.86, 0.028\\ \hline
Na\(_2\) &    278.20 &  1.595 &  0.4275 &  7.33e6 & 0.81, 0.012\\ \hline
K\(_2\)  &    359.65 &  1.381 &  0.3273 &  7.30e7 & 0.85, 0.012\\ \hline
Ar\(_2\) &   4994.79 &  2.921 &  0.2959 &  3.12e7 & 0.70, 0.012\\ \hline
Kr\(_2\) &   9610.07 &  2.805 &  0.2759 &  6.23e7 & 0.90, 0.0004\\ \hline
\end{tabular}
\end{ruledtabular}
\end{table}
\par
\par
\begin{figure}
\centering
\input {RKRO2.tex}
\caption{The hybrid form \(\tcr{V}\) 
with the calculated coefficients 
(\(a\), \(b\), \(c\), \& \(d\), Table~\ref{tab:table3})
(solid, red) fits
the RKR spectral points for 
the ground state of
molecular oxygen (diamonds, cyan)
nearly as well as does the hybrid form \(\tcnb{V_f}\) 
with the fitted coefficients 
(\(a\), \(b\), \(c\), \& \(d\), Table~\ref{tab:table2})
(dashes, dark blue)\@.}
\label{rkrO2}
\end{figure}
\par
\begin{figure}
\centering
\input {RKRNO.tex}
\caption{The hybrid form \(\tcr{V}\) 
with the calculated coefficients 
(\(a\), \(b\), \(c\), \& \(d\), Table~\ref{tab:table3})
(solid, red) fits
the RKR spectral points for 
the ground state of
nitric oxide (diamonds, cyan)
nearly as well as does the hybrid form \(\tcnb{V_f}\) 
with the fitted coefficients 
(\(a\), \(b\), \(c\), \& \(d\), Table~\ref{tab:table2})
(dashes, dark blue)\@.}
\label{rkrNO}
\end{figure}
\par
\par
\begin{figure}
\centering
\input {RKRI2.tex}
\caption{The hybrid form \(\tcr{V}\) 
with the calculated coefficients 
(\(a\), \(b\), \(c\), \& \(d\), Table~\ref{tab:table3})
(solid, red) fits
the RKR spectral points for 
the ground state of
molecular iodine (diamonds, cyan)
nearly as well as does the hybrid form \(\tcnb{V_f}\) 
with the fitted coefficients 
(\(a\), \(b\), \(c\), \& \(d\), Table~\ref{tab:table2})
(dashes, dark blue)\@.}
\label{rkrI2}
\end{figure}
\par
\begin{figure}
\centering
\input {RKRLi2.tex}
\caption{The hybrid form \(\tcr{V}\) 
with the calculated coefficients 
(\(a\), \(b\), \(c\), \& \(d\), Table~\ref{tab:table3})
(solid, red) fits
the RKR spectral points for 
the ground state of the
lithium dimer (diamonds, cyan)
nearly as well as does the hybrid form \(\tcnb{V_f}\) 
with the fitted coefficients 
(\(a\), \(b\), \(c\), \& \(d\), Table~\ref{tab:table2})
(dashes, dark blue)\@.}
\label{rkrLi2}
\end{figure}
\par
\begin{figure}
\centering
\input {RKRNa2.tex}
\caption{The hybrid form \(\tcr{V}\) 
with the calculated coefficients 
(\(a\), \(b\), \(c\), \& \(d\), Table~\ref{tab:table3})
(solid, red) fits
the RKR spectral points for 
the ground state of the
sodium dimer (diamonds, cyan) well, but not quite 
as well as does the hybrid form \(\tcnb{V_f}\) 
with the fitted coefficients 
(\(a\), \(b\), \(c\), \& \(d\), Table~\ref{tab:table2})
(dashes, dark blue)\@.}
\label{rkrNa2}
\end{figure}
\par
\begin{figure}
\centering
\input {RKRK2.tex}
\caption{The hybrid form \(\tcr{V}\) 
with the calculated coefficients 
(\(a\), \(b\), \(c\), \& \(d\), Table~\ref{tab:table3})
(solid, red) fits
the RKR spectral points for 
the ground state of the
potassium dimer (diamonds, cyan) well, but not quite
as well as does the hybrid form \(\tcnb{V_f}\) 
with the fitted coefficients 
(\(a\), \(b\), \(c\), \& \(d\), Table~\ref{tab:table2})
(dashes, dark blue)\@.}
\label{rkrK2}
\end{figure}
\par
\par
\begin{figure}
\centering
\input {RKRKr2.tex}
\caption{The hybrid form \(\tcr{V}\) 
with the calculated coefficients 
(\(a\), \(b\), \(c\), \& \(d\), Table~\ref{tab:table3})
(solid, red) fits
the RKR spectral points for 
the ground state of 
the krypton dimer (diamonds, cyan)
nearly as well as does the hybrid form \(\tcnb{V_f}\) 
with the fitted coefficients 
(\(a\), \(b\), \(c\), \& \(d\), Table~\ref{tab:table2})
(dashes, dark blue)\@.}
\label{rkrKr2}
\end{figure}

\par
The fitted values of the parameter \(d\) 
in Table~\ref{tab:table2} are 
less well defined than those of \(a\), \(b\), and \(c\)\@.
The empirical rule of thumb
\beq
d = 7.1 \, \mbox{\AA}^{12} 
+ 2.89 \, \frac{C_6^3}{E_0^3 \, r_0^6}
+ 0.468 \, \frac{E_0}{\mbox{eV}} \, \frac{r_0^{13}}{\mbox{\AA}} 
\label{d emp}
\eeq
gives the 11 values of \(d\) listed
in Table~\ref{tab:table3}
and roughly approximates those
of Table~\ref{tab:table2}\@.
\section{Three Formulas\label{abc}}
Suppose we use the rule (\ref{d emp})
for the fudge-factor \(d\)
and take the London coefficient
\(C_6\) from Table~\ref{tab:table1}\@.
How do we find the parameters \(a\),
\(b\), and \(c\) that make the
hybrid potential \(V(r)\) have its minimum
at \(r = r_0\) with \(V(r_0) = - E_0\) 
and with curvature \(V^{''}(r_0) = k\)?
\par
Let us write the hybrid form 
as the sum 
\beq
V(r) = v(r) + w(r)
\label {v+w}
\eeq
of the \(a, b, c\) terms \(v(r)\)
\beq
v(r) = a \, e^{-br} ( 1 - c \, r )
\label {v}
\eeq
and the \(C_6, d\) terms \(w(r)\)
\beq
w(r) = - \frac{C_6}{r^6 + d \, r^{-6}}.
\label {w}
\eeq
Since \(C_6\) and \(d\) are given,
the function \(w(r)\)
and its derivatives
\beq
w'(r) = 6 \, C_6 \, r^5 \, \frac{r^{12} - d}{(r^{12} + d)^2}
\label {w'}
\eeq 
and 
\beq
w^{\prime\prime}(r) = - 6 \, C_6 \, r^4 \, 
\frac{( 7 r^{12} - d )(r^{12} - 5d)}
{(r^{12} + d)^3}
\label {w''}
\eeq
are determined.  
The condition that
\(V(r_0) = - E_0\) then is
\beq
a \, e^{-br_0}(1 - c \, r_0) = - w(r_0) - E_0 
\label {a=}
\eeq
which implies that \(a\) is
\beq
a = - (w(r_0) + E_0) \, e^{br_0} / (1 - c \, r_0).
\label {ais}
\eeq
The condition that \(V'(r_0) = 0\) is
\beq
a \, e^{-br_0}(b + c - b \, c \, r_0) = w'(r_0)
\label {b=}
\eeq
which together with (\ref{ais}) gives \(b\) as
\beq
b = - c/(1 - c \, r_0) - w'(r_0)/(w(r_0) + E_0).
\label {bis}
\eeq
Finally, the condition \(V^{''}(r_0) = k\) is
\beq
a \, b e^{-br_0}(b - b \, c \, r_0 + 2 \, c) = k - w^{\prime\prime}(r_0)
\label {c=}
\eeq
which with (\ref{ais}) for \(a\) and (\ref{bis}) for \(b\)
is a quadratic equation for \(c\)
with roots
\beq
c = \left[
r_0 \pm \sqrt{ \frac{(w(r_0) + E_0)^2}
{w'(r_0)^2 + 
(k - w^{\prime\prime}(r_0))(w(r_0) + E_0)} }
\right]^{-1} .
\label {cis}
\eeq
The minus sign implies
\beq
c > 1/r_0 \Longleftrightarrow 1 - c \, r_0 < 0
\label {minus sign}
\eeq
while the plus sign implies
\beq
c < 1/r_0 \Longleftrightarrow 1 - c \, r_0 > 0.
\label {plus sign}
\eeq
\par
A covalently bonded pair 
will have \(v(r_0) < 0\),
which implies \(1 - c r_0 < 0\)
and so \(c > 1/r_0\),
which entails the minus sign.
One may choose the minus sign 
for the 11 pairs of neutral atoms 
considered in this paper.
A further discussion
of the choice of sign
appears in the Appendix.
\par
If one uses the rule of thumb (\ref{d emp})
for \(d\) and the values of 
\(r_0\), \(E_0\), \(k\), \& \(C_6\)
from Table~\ref{tab:table1}, 
then one may find \(c\)
from condition (\ref{cis}) with a choice of sign,
and then \(b\) from (\ref{bis}),
and then \(a\) from (\ref{ais})\@.
The resulting values of \(a\), \(b\), \(c\), \& \(d\)
for the 11 pairs of neutral atoms
H\(_2\), N\(_2\), O\(_2\), NO, OH, I\(_2\), 
Li\(_2\), Na\(_2\), K\(_2\),
Ar--Ar, \& Kr--Kr
are listed in Table~\ref{tab:table3}
for the case of a minus sign in Eq.~(\ref{cis});
also shown are the rms errors \(\Delta V(r)\) for
\(r \ge f r_0\) for the \(f\)'s of 
Table~\ref{tab:table2}\@.
Figures~\ref{rkrH2}--\ref{rkrKr2} show 
the hybrid form \(\tcr{V}\)
with these values of \(a\), \(b\), \(c\), \& \(d\)
and with the \(C_6\)'s of Table~\ref{tab:table1}
as solid red curves.  They closely
trace the empirical
potentials (diamonds, cyan) and 
the fitted 
hybrid forms \(\tcnb{V_f}\) (dashes, dark blue)\@.

\section{Conclusions\label{conclusions}}

By using Eqs.~(\ref{d emp}, \ref{ais}, 
\ref{bis}, \ref{cis}, \& \ref{minus sign}),
one may build an accurate hybrid potential~(\ref{hybrid})
for the ground state of a pair of neutral atoms
from their internuclear separation,
the depth and curvature 
of their potential at its minimum,
and from their 
van der Waals coefficient \(C_6\)\@.
The hybrid potential therefore is applicable
to pairs of neutral atoms for which
no empirical potential is available.
Given the differences between it
and the 
Lennard-Jones, harmonic, Morse,
Varnshi, and Hulburt-Hirschfelder potentials, 
it would be worthwhile
to examine the consequences of
these differences in Monte Carlo searches
for low-energy states of biomolecules
and in numerical simulations
of phase transitions and reactions
far from equilibrium.
\begin{acknowledgments}
Thanks to G.~Groenenboom for sending me
a copy of the potential of~\cite{Groenenboom2007}
and to S.~Atlas, S.~T.~P. Boyd, H.-B. Broeker, D.~Cromer,
K.~Dill, M.~Fleharty, S.~C. Foster, R.~J. Le Roy,
W.~J. Meath, V.~A.~Parsegian, R.~Pastor, 
R.~Podgornik, E.~Riley, 
B.~Rivers, D.~Sergatskov, J.~Thomas, S.~Valone
and W.~Zemke
for advice.
\end{acknowledgments}
\appendix*
\section{The Choice of Sign}
\par
The choice of the minus sign 
in Eq.(\ref{cis}) implies 
that the fudge factor \(d\) 
must exceed a certain lower limit.
The parameter \(a\) is the value
of the potential \(V(r)\) at \(r = 0\)
and so must be positive.
The minus-sign inequalities (\ref{minus sign})
imply that \(1 - c r_0 < 0\)\@.
Thus Eq.(\ref{ais}) 
will give a positive value for \(a\) 
only if 
\beq
w(r_0) + E_0 > 0
\label {need a>0}
\eeq
which by (\ref{w}) implies 
the lower limit
\beq
d > \frac{C_6 \, r_0^6}{E_0} - r_0^{12}
\label {lower limit on d}
\eeq
on the fudge factor \(d\)\@.
The values of \(d\) listed in
Tables~\ref{tab:table2} \& \ref{tab:table3}
satisfy this constraint.
\par
The choice of the plus sign 
in Eq.(\ref{cis}) implies two upper limits
on the fudge factor \(d\)\@.
The plus-sign inequalities (\ref{plus sign})
imply that \(1 - c r_0 > 0\)\@.
Thus Eq.(\ref{ais}) 
will give a positive value for \(a\) 
only if 
\beq
w(r_0) + E_0 < 0
\label {need a<0}
\eeq
which by (\ref{w}) implies 
the upper limit
\beq
d < \frac{C_6 \, r_0^6}{E_0} - r_0^{12}
\label {1st upper limit on d}
\eeq
on the fudge factor \(d\)\@.
The right-hand-side of this
inequality is negative for the pairs
N\(_2\), O\(_2\), NO, \& I\(_2\)
and less than unity for the
pairs H\(_2\) \& OH,
as may be seen from the values
of \(C_6\), \(r_0\), \& \(E_0\)
listed in Table~\ref{tab:table1}\@.
Thus the minus sign is required for the pairs
N\(_2\), O\(_2\), NO, \& I\(_2\)
and strongly indicated for the
pairs H\(_2\) \& OH\@.
\par
The second upper limit on \(d\) arises
because both \(b\) and \(c\) must be positive
(\(b > 0\) because \(v(r)\) must vanish as \(r \to \infty\),
and \(c>0\) because \(c < 0 \) implies \(v(r)>0\))\@.
Thus if the positive sign is chosen
in the formula (\ref{cis}) for \(c\),
then \(1 - c r_0 > 0\), and also
\(w(r_0) + E_0 < 0\) by (\ref{need a<0}),
and so Eq.(\ref{bis}) will give 
a positive value for \(b\) only if 
\(w'(r_0) > 0\), which by (\ref{w'})
implies the upper limit
\beq
d < r_0^{12}.
\label {2d upper limit on d}
\eeq
\par
The curves displayed in Figs.~\ref{rkrH2}--\ref{rkrKr2}
all correspond to the choice of a minus sign in Eq.(\ref{cis})\@.
But for the five
pairs Li\(_2\), Na\(_2\), K\(_2\), Ar--Ar, \& Kr--Kr,
one also may get good fits to the empirical data
by using the plus sign (\ref{plus sign})\@.
The resulting values of \(a\), \(b\), \(c\), \& \(d\)
appear in Table~\ref{tab:table4}\@.
The \(d\)'s obey (\ref{1st upper limit on d} \& \ref{2d upper limit on d})\@.
\begin{table}
\caption{\label{tab:table4}The values of the coefficients
\(a\), \(b\), \(c\), \& \(d\) that best fit 
the hybrid potential \(V(r)\) (Eq.~(\ref{hybrid})) 
to RKR data for Li\(_2\), Na\(_2\), \& K\(_2\)
and to empirical potentials for Ar--Ar \& Kr--Kr 
while respecting Eqs.~(\ref{ais}, \ref{bis}, \& \ref{cis}) 
with a plus sign in (\ref{cis})\@.
\(\Delta V(r)\) is the rms error in \(V\) for \(r \ge f \, r_0\)\@.}
\begin{ruledtabular}
\begin{tabular}{|c|c|c|c|c|c|}  
\  & \(a\) (eV) & \(b\) (\AA\(^{-1}\)) 
& \(c\) (\AA\(^{-1}\)) & \(d\) (\AA\(^{12}\)) & f,\,\, \(\Delta V\!\) (eV)\\
\hline 
Li\(_2\)&  1136.21 & 1.8218 & 0.3225 &    869. & 0.77, 0.037\\ \hline
Na\(_2\)&   648.22 & 1.5692 & 0.3034 &   2692. & 0.90, 0.004\\ \hline
K\(_2\) &   500.31 & 1.2673 & 0.2486 &  33770. & 0.70, 0.012\\ \hline
Ar\(_2\)&  1754.01 & 2.7054 & 0.2620 &  1.57e5 & 0.70, 0.001\\ \hline
Kr\(_2\)& 2725.43  & 2.5622 & 0.2485 &   1.0e6 & 0.90, 0.0003\\ \hline
\end{tabular}
\end{ruledtabular}
\end{table}
\bibliography{chem,cs,physics,vdw,bio,proteins,math,biochem}

\begin{thebibliography}{29}
\expandafter\ifx\csname natexlab\endcsname\relax\def\natexlab#1{#1}\fi
\expandafter\ifx\csname bibnamefont\endcsname\relax
  \def\bibnamefont#1{#1}\fi
\expandafter\ifx\csname bibfnamefont\endcsname\relax
  \def\bibfnamefont#1{#1}\fi
\expandafter\ifx\csname citenamefont\endcsname\relax
  \def\citenamefont#1{#1}\fi
\expandafter\ifx\csname url\endcsname\relax
  \def\url#1{\texttt{#1}}\fi
\expandafter\ifx\csname urlprefix\endcsname\relax\def\urlprefix{URL }\fi
\providecommand{\bibinfo}[2]{#2}
\providecommand{\eprint}[2][]{\url{#2}}

\bibitem[{\citenamefont{Lennard-Jones}(1931)}]{Lennard-Jones1931}
\bibinfo{author}{\bibfnamefont{J.~E.} \bibnamefont{Lennard-Jones}},
  \bibinfo{journal}{\textsl{Proc.\ Phys.\ Soc.}} \textbf{\bibinfo{volume}{43}},
  \bibinfo{pages}{461} (\bibinfo{year}{1931}).

\bibitem[{\citenamefont{Cahill and Parsegian}(2004)}]{CahillPar2004}
\bibinfo{author}{\bibfnamefont{K.}~\bibnamefont{Cahill}} \bibnamefont{and}
  \bibinfo{author}{\bibfnamefont{V.~A.} \bibnamefont{Parsegian}},
  \bibinfo{journal}{\textsl{J.~ Chem. Phys.}} \textbf{\bibinfo{volume}{121}},
  \bibinfo{pages}{10839} (\bibinfo{year}{2004}),
  \bibinfo{note}{\href{http://arxiv.org/abs/q-bio.BM/0410018}
  {arxiv.org/q-bio.BM/0410018}}.

\bibitem[{\citenamefont{Ren and J.~W.~Ponder}(2003)}]{Ponder2003}
\bibinfo{author}{\bibfnamefont{P.}~\bibnamefont{Ren}} \bibnamefont{and}
  \bibinfo{author}{\bibfnamefont{J.~W.} \bibnamefont{J.~W.~Ponder}},
  \bibinfo{journal}{\textsl{J. Phys. Chem. B}} \textbf{\bibinfo{volume}{107}},
  \bibinfo{pages}{5933} (\bibinfo{year}{2003}).

\bibitem[{\citenamefont{Pearlman et~al.}(1995)\citenamefont{Pearlman, Case,
  Caldwell, Ross, Cheatham~III, DeBolt, Ferguson, Seibel, and
  Kollman}}]{Amber71995}
\bibinfo{author}{\bibfnamefont{D.~A.} \bibnamefont{Pearlman}},
  \bibinfo{author}{\bibfnamefont{D.~A.} \bibnamefont{Case}},
  \bibinfo{author}{\bibfnamefont{J.~W.} \bibnamefont{Caldwell}},
  \bibinfo{author}{\bibfnamefont{W.~S.} \bibnamefont{Ross}},
  \bibinfo{author}{\bibfnamefont{T.~E.} \bibnamefont{Cheatham~III}},
  \bibinfo{author}{\bibfnamefont{S.}~\bibnamefont{DeBolt}},
  \bibinfo{author}{\bibfnamefont{D.}~\bibnamefont{Ferguson}},
  \bibinfo{author}{\bibfnamefont{G.}~\bibnamefont{Seibel}}, \bibnamefont{and}
  \bibinfo{author}{\bibfnamefont{P.}~\bibnamefont{Kollman}},
  \bibinfo{journal}{Comp.\ Phys.\ Commun.} \textbf{\bibinfo{volume}{91}},
  \bibinfo{pages}{1} (\bibinfo{year}{1995}).

\bibitem[{\citenamefont{Cahill and Parsegian}(2003)}]{CahillPar2003}
\bibinfo{author}{\bibfnamefont{K.}~\bibnamefont{Cahill}} \bibnamefont{and}
  \bibinfo{author}{\bibfnamefont{V.~A.} \bibnamefont{Parsegian}}
  (\bibinfo{year}{2003}),
  \bibinfo{note}{\href{http://arxiv.org/abs/q-bio.BM/0312005}
  {q-bio.BM/0312005}}.

\bibitem[{\citenamefont{Morse}(1929)}]{Morse1929}
\bibinfo{author}{\bibfnamefont{P.~M.} \bibnamefont{Morse}},
  \bibinfo{journal}{\textsl{Phys.\ Rev.}} \textbf{\bibinfo{volume}{34}},
  \bibinfo{pages}{57} (\bibinfo{year}{1929}).

\bibitem[{\citenamefont{Varnshi}(1957)}]{Varnshi1957}
\bibinfo{author}{\bibfnamefont{Y.~P.} \bibnamefont{Varnshi}},
  \bibinfo{journal}{\textsl{Rev.\ Mod.\ Phys.}} \textbf{\bibinfo{volume}{29}},
  \bibinfo{pages}{664} (\bibinfo{year}{1957}).

\bibitem[{\citenamefont{Hulbert and Hirschfelder}(1961)}]{Hulbert1961}
\bibinfo{author}{\bibfnamefont{H.~M.} \bibnamefont{Hulbert}} \bibnamefont{and}
  \bibinfo{author}{\bibfnamefont{J.~O.} \bibnamefont{Hirschfelder}},
  \bibinfo{journal}{\textsl{J.\ Chem.\ Phys.}} \textbf{\bibinfo{volume}{35}},
  \bibinfo{pages}{1901(L)} (\bibinfo{year}{1961}).

\bibitem[{\citenamefont{Steele et~al.}(1962)\citenamefont{Steele, Lippincott,
  and Vanderslice}}]{Vanderslice1962}
\bibinfo{author}{\bibfnamefont{D.}~\bibnamefont{Steele}},
  \bibinfo{author}{\bibfnamefont{E.~R.} \bibnamefont{Lippincott}},
  \bibnamefont{and} \bibinfo{author}{\bibfnamefont{J.~T.}
  \bibnamefont{Vanderslice}}, \bibinfo{journal}{\textsl{Rev.\ Mod.\ Phys.}}
  \textbf{\bibinfo{volume}{34}}, \bibinfo{pages}{239} (\bibinfo{year}{1962}).

\bibitem[{\citenamefont{Krupenie}(1972)}]{Krupenie1972}
\bibinfo{author}{\bibfnamefont{P.~H.} \bibnamefont{Krupenie}},
  \bibinfo{journal}{\textsl{J.\ Phys.\ Chem.\ Ref.\ Data}}
  \textbf{\bibinfo{volume}{1}}, \bibinfo{pages}{423} (\bibinfo{year}{1972}).

\bibitem[{\citenamefont{Mitroy and Bromley}(2005)}]{Bromley2005}
\bibinfo{author}{\bibfnamefont{J.}~\bibnamefont{Mitroy}} \bibnamefont{and}
  \bibinfo{author}{\bibfnamefont{M.~W.~J.} \bibnamefont{Bromley}},
  \bibinfo{journal}{Phys. Rev. A} \textbf{\bibinfo{volume}{71}},
  \bibinfo{pages}{032709} (\bibinfo{year}{2005}).

\bibitem[{\citenamefont{Zeiss and Meath}(1977)}]{Zeiss&Meath1977}
\bibinfo{author}{\bibfnamefont{G.~D.} \bibnamefont{Zeiss}} \bibnamefont{and}
  \bibinfo{author}{\bibfnamefont{W.~J.} \bibnamefont{Meath}},
  \bibinfo{journal}{\textsl{Mol.\ Phys.}} \textbf{\bibinfo{volume}{33}},
  \bibinfo{pages}{1155} (\bibinfo{year}{1977}).

\bibitem[{\citenamefont{Varandas and Voronin}(1995)}]{Varandas1995}
\bibinfo{author}{\bibfnamefont{A.~J.~C.} \bibnamefont{Varandas}}
  \bibnamefont{and} \bibinfo{author}{\bibfnamefont{A.~I.}
  \bibnamefont{Voronin}}, \bibinfo{journal}{Chem. Phys.}
  \textbf{\bibinfo{volume}{194(1)}}, \bibinfo{pages}{91}
  (\bibinfo{year}{1995}).

\bibitem[{\citenamefont{Chu and Dalgarno}(2004)}]{Dalgarno2004}
\bibinfo{author}{\bibfnamefont{X.}~\bibnamefont{Chu}} \bibnamefont{and}
  \bibinfo{author}{\bibfnamefont{A.}~\bibnamefont{Dalgarno}},
  \bibinfo{journal}{\textsl{J.\ Chem.\ Phys.}}
  \textbf{\bibinfo{volume}{121(9)}}, \bibinfo{pages}{4083}
  (\bibinfo{year}{2004}).

\bibitem[{\citenamefont{Aziz}(1993)}]{Aziz1993}
\bibinfo{author}{\bibfnamefont{R.~A.} \bibnamefont{Aziz}},
  \bibinfo{journal}{\textsl{J.\ Chem.\ Phys.}} \textbf{\bibinfo{volume}{99}},
  \bibinfo{pages}{4518} (\bibinfo{year}{1993}).

\bibitem[{\citenamefont{Dham et~al.}(1989)\citenamefont{Dham, Allnatt, Meath,
  and Aziz}}]{Dham1989}
\bibinfo{author}{\bibfnamefont{A.~K.} \bibnamefont{Dham}},
  \bibinfo{author}{\bibfnamefont{A.~R.} \bibnamefont{Allnatt}},
  \bibinfo{author}{\bibfnamefont{W.~J.} \bibnamefont{Meath}}, \bibnamefont{and}
  \bibinfo{author}{\bibfnamefont{R.~A.} \bibnamefont{Aziz}},
  \bibinfo{journal}{\textsl{Mol.\ Phys.}} \textbf{\bibinfo{volume}{67}},
  \bibinfo{pages}{1291} (\bibinfo{year}{1989}).

\bibitem[{\citenamefont{Le~Roy et~al.}(2006)\citenamefont{Le~Roy, Yiye~Huang,
  and Jary}}]{Jary2006}
\bibinfo{author}{\bibfnamefont{R.~J.} \bibnamefont{Le~Roy}},
  \bibinfo{author}{\bibfnamefont{Y.}~\bibnamefont{Yiye~Huang}},
  \bibnamefont{and} \bibinfo{author}{\bibfnamefont{C.}~\bibnamefont{Jary}},
  \bibinfo{journal}{\textsl{J.\ Chem.\ Phys.}}
  \textbf{\bibinfo{volume}{125(16)}}, \bibinfo{pages}{164309}
  (\bibinfo{year}{2006}).

\bibitem[{\citenamefont{Coxon and Melville}(2006)}]{Melville2006}
\bibinfo{author}{\bibfnamefont{J.~A.} \bibnamefont{Coxon}} \bibnamefont{and}
  \bibinfo{author}{\bibfnamefont{T.~C.} \bibnamefont{Melville}},
  \bibinfo{journal}{\textsl{J. Mol. Spectrosc.}}
  \textbf{\bibinfo{volume}{235(2)}}, \bibinfo{pages}{235}
  (\bibinfo{year}{2006}).

\bibitem[{\citenamefont{Samuelis et~al.}(2000)\citenamefont{Samuelis, Tiesinga,
  Laue, Elbs, Kn{\"{o}}ckel, and Tiemann}}]{Tiemann2000}
\bibinfo{author}{\bibfnamefont{C.}~\bibnamefont{Samuelis}},
  \bibinfo{author}{\bibfnamefont{E.}~\bibnamefont{Tiesinga}},
  \bibinfo{author}{\bibfnamefont{T.}~\bibnamefont{Laue}},
  \bibinfo{author}{\bibfnamefont{M.}~\bibnamefont{Elbs}},
  \bibinfo{author}{\bibfnamefont{H.}~\bibnamefont{Kn{\"{o}}ckel}},
  \bibnamefont{and} \bibinfo{author}{\bibfnamefont{E.}~\bibnamefont{Tiemann}},
  \bibinfo{journal}{Phys. Rev. A} \textbf{\bibinfo{volume}{63(1)}},
  \bibinfo{pages}{012710} (\bibinfo{year}{2000}).

\bibitem[{\citenamefont{Derevianko et~al.}(1999)\citenamefont{Derevianko,
  Johnson, Safronova, and Babb}}]{Johnson1999}
\bibinfo{author}{\bibfnamefont{A.}~\bibnamefont{Derevianko}},
  \bibinfo{author}{\bibfnamefont{W.~R.} \bibnamefont{Johnson}},
  \bibinfo{author}{\bibfnamefont{M.~S.} \bibnamefont{Safronova}},
  \bibnamefont{and} \bibinfo{author}{\bibfnamefont{J.~F.} \bibnamefont{Babb}},
  \bibinfo{journal}{Phys. Rev. Lett.} \textbf{\bibinfo{volume}{82(18)}},
  \bibinfo{pages}{3589} (\bibinfo{year}{1999}).

\bibitem[{\citenamefont{Amiot et~al.}(1995)\citenamefont{Amiot, Verg{\`{e}}s,
  and Fellows}}]{Amiot1995}
\bibinfo{author}{\bibfnamefont{C.}~\bibnamefont{Amiot}},
  \bibinfo{author}{\bibfnamefont{J.}~\bibnamefont{Verg{\`{e}}s}},
  \bibnamefont{and} \bibinfo{author}{\bibfnamefont{C.~E.}
  \bibnamefont{Fellows}}, \bibinfo{journal}{J. Chem. Phys.}
  \textbf{\bibinfo{volume}{103(9)}}, \bibinfo{pages}{3350}
  (\bibinfo{year}{1995}).

\bibitem[{\citenamefont{Amiot}(1991)}]{Amiot1991}
\bibinfo{author}{\bibfnamefont{C.}~\bibnamefont{Amiot}},
  \bibinfo{journal}{\textsl{J. Mol. Spectrosc.}}
  \textbf{\bibinfo{volume}{147}}, \bibinfo{pages}{370} (\bibinfo{year}{1991}).

\bibitem[{\citenamefont{Weissman et~al.}(1963)\citenamefont{Weissman,
  Vanderslice, and Battino}}]{Weissman1963}
\bibinfo{author}{\bibfnamefont{S.}~\bibnamefont{Weissman}},
  \bibinfo{author}{\bibfnamefont{J.~T.} \bibnamefont{Vanderslice}},
  \bibnamefont{and} \bibinfo{author}{\bibfnamefont{R.}~\bibnamefont{Battino}},
  \bibinfo{journal}{\textsl{J.\ Chem.\ Phys.}} \textbf{\bibinfo{volume}{39}},
  \bibinfo{pages}{2226} (\bibinfo{year}{1963}).

\bibitem[{\citenamefont{Lofthus and Krupenie}(1977)}]{Krupenie1977}
\bibinfo{author}{\bibfnamefont{A.}~\bibnamefont{Lofthus}} \bibnamefont{and}
  \bibinfo{author}{\bibfnamefont{P.~H.} \bibnamefont{Krupenie}},
  \bibinfo{journal}{\textsl{J.\ Phys.\ Chem.\ Ref.\ Data}}
  \textbf{\bibinfo{volume}{6}}, \bibinfo{pages}{113} (\bibinfo{year}{1977}).

\bibitem[{\citenamefont{Rydberg}(1931)}]{Rydberg1931}
\bibinfo{author}{\bibfnamefont{R.}~\bibnamefont{Rydberg}},
  \bibinfo{journal}{\textsl{Z.\ Phys.}} \textbf{\bibinfo{volume}{73}},
  \bibinfo{pages}{376} (\bibinfo{year}{1931}).

\bibitem[{\citenamefont{Klein}(1932)}]{Klein1932}
\bibinfo{author}{\bibfnamefont{O.}~\bibnamefont{Klein}},
  \bibinfo{journal}{\textsl{Z.\ Phys.}} \textbf{\bibinfo{volume}{76}},
  \bibinfo{pages}{226} (\bibinfo{year}{1932}).

\bibitem[{\citenamefont{Rees}(1947)}]{Rees1947}
\bibinfo{author}{\bibfnamefont{A.~L.~G.} \bibnamefont{Rees}},
  \bibinfo{journal}{\textsl{Proc.\ Phys.\ Soc.\ (London)}}
  \textbf{\bibinfo{volume}{59}}, \bibinfo{pages}{998} (\bibinfo{year}{1947}).

\bibitem[{\citenamefont{van~der Loo and Groenenboom}(2007)}]{Groenenboom2007}
\bibinfo{author}{\bibfnamefont{M.~P.~J.} \bibnamefont{van~der Loo}}
  \bibnamefont{and} \bibinfo{author}{\bibfnamefont{G.~C.}
  \bibnamefont{Groenenboom}}, \bibinfo{journal}{\textsl{J. Chem. Phys.}}
  \textbf{\bibinfo{volume}{126}}, \bibinfo{pages}{114314}
  (\bibinfo{year}{2007}).

\bibitem[{\citenamefont{C\^{o}t\'e et~al.}(1998)\citenamefont{C\^{o}t\'e,
  Dalgarno, Wang, and Stwalley}}]{Stwalley1998}
\bibinfo{author}{\bibfnamefont{R.}~\bibnamefont{C\^{o}t\'e}},
  \bibinfo{author}{\bibfnamefont{A.}~\bibnamefont{Dalgarno}},
  \bibinfo{author}{\bibfnamefont{H.}~\bibnamefont{Wang}}, \bibnamefont{and}
  \bibinfo{author}{\bibfnamefont{W.~C.} \bibnamefont{Stwalley}},
  \bibinfo{journal}{\textsl{Phys.\ Rev.\ A}} \textbf{\bibinfo{volume}{57}},
  \bibinfo{pages}{R4118} (\bibinfo{year}{1998}).

\end{thebibliography}
\end{document}